\documentclass[useAMS,usenatbib]{mn2e}
\usepackage{epsfig}
\usepackage{multirow}

\title[Constraints on kHz QPO models and ...]
{Constraints on kHz QPO models and stellar EOSs from SAX J1808.4-3658, Cyg X-2 and 4U 1820-30}

\author[D. H. Wang et al.]
{D. H. Wang$^{1,2}$, L. Chen$^1$, C. M. Zhang$^2\thanks{zhangcm@bao.ac.cn, huazai05105220@163.com}$, Y.J. Lei$^2$, J. L. Qu$^3$
\\
$^1$Astronomy Department, Beijing Normal University, Beijing, 100875, China\\
$^2$National Astronomical Observatories, University of Chinese Academy of Sciences, Beijing, 100012, China\\
$^3$Institute of High Energy Physics, University of Chinese Academy of Sciences, Beijing, 100049, China\\
}
\date{Released 2002 Xxxxx XX}

\pagerange{\pageref{firstpage}--\pageref{lastpage}} \pubyear{2002}

\begin{document}

\maketitle

\label{firstpage}

\begin{abstract}

We test the relativistic precession model (RPM) and the MHD Alfv\'en wave oscillation model (AWOM) {\bf for the kHz QPOs } by the sources
 with measured NS masses and twin kHz QPO frequencies.  For RPM, the derived NS mass of Cyg X-2 (SAX J1808.4-3658 and 4U 1820-30) is $1.96\pm0.10\,M_\odot$ ($2.83\pm0.04\,M_\odot$ and $1.85\pm0.02\,M_\odot$), which is $\sim30\%$ (100\% and 40\%) higher than the measured result  $1.5\pm0.3\,M_\odot$ ($<1.4\,M_\odot$ and $1.29^{+0.19}_{-0.07}\,M_\odot$).
For AWOM, where the free parameter of model is the density of star, we infer the NS radii to be around $10\sim20$\,km for the above
  three sources, based on which we can  infer the matter compositions inside NSs with the help of the equations of state (EOSs).
  In particular, for SAX J1808.4-3658, AWOM shows a lower mass density of its NS than those of the other known kHz QPO sources,  with the  radius range of  $17-20\,\rm km$, which excludes the
   strange quark matter inside its star.

\end{abstract}


\section{Introduction}
\begin{figure*}
\includegraphics[width=8cm]{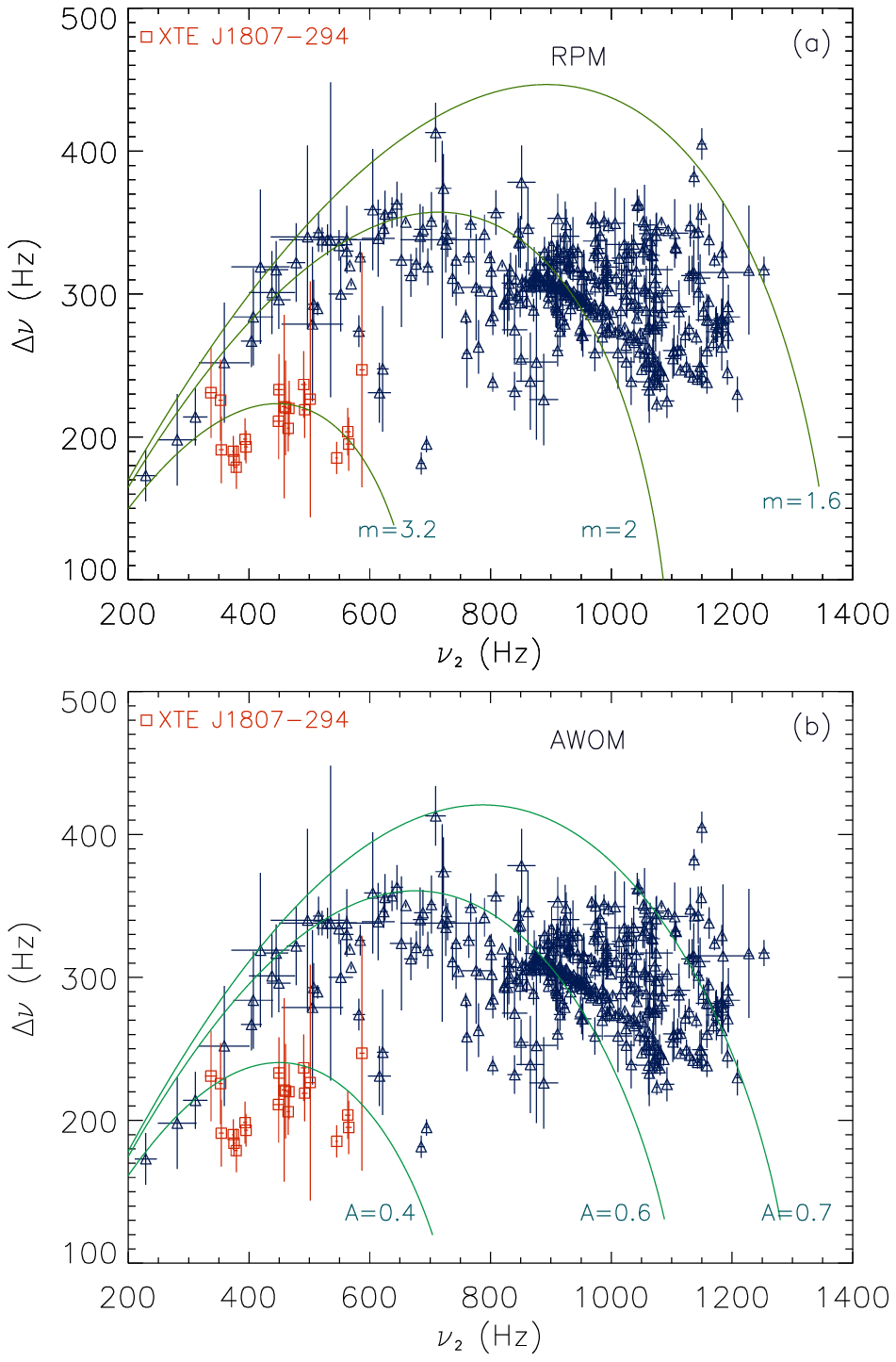}
\includegraphics[width=8cm]{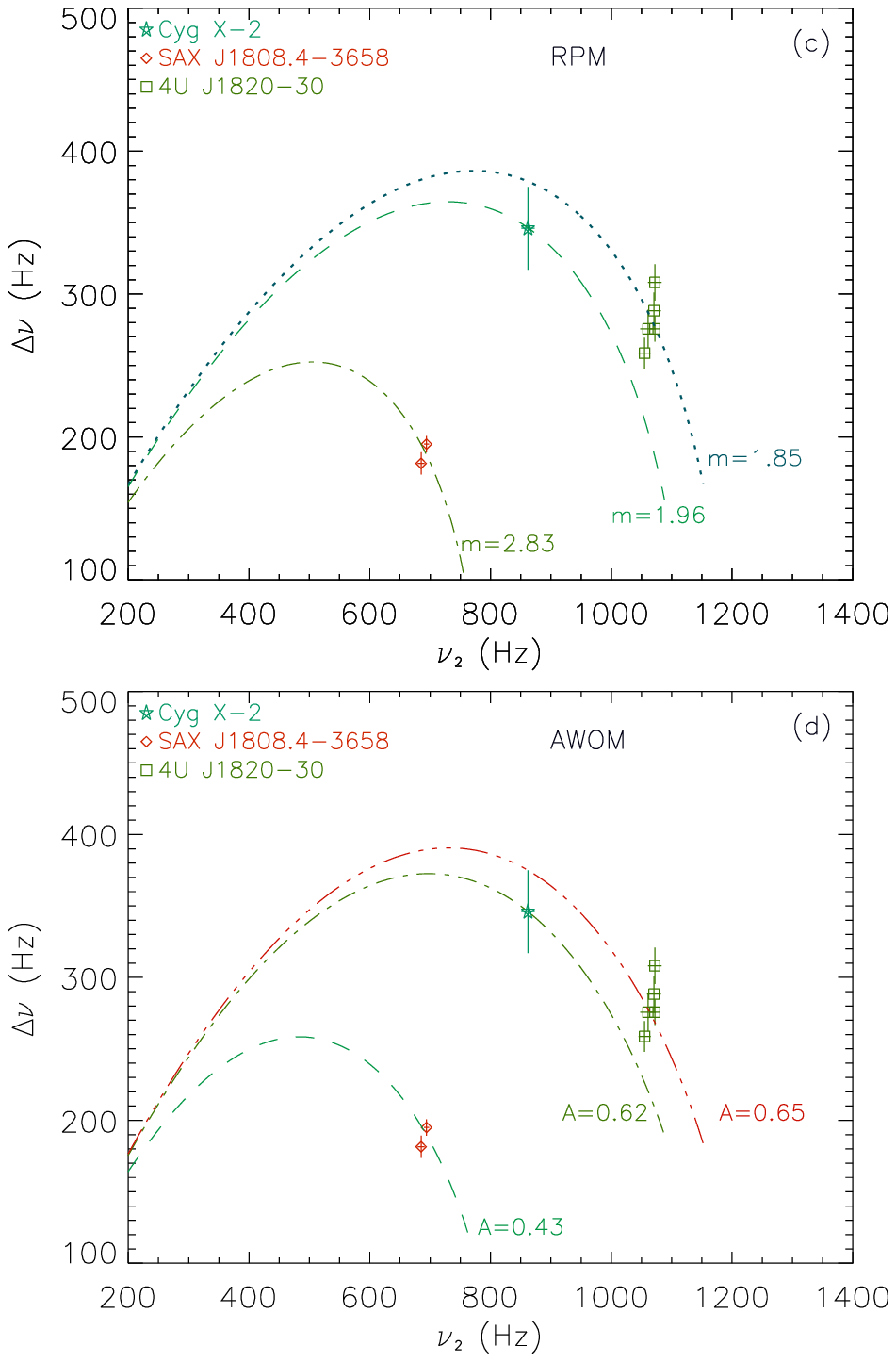}
\caption{(a) [(b)] shows the plot of $\Delta\nu~vs.~\nu_2$, including 393 pairs of kHz QPOs of 26 sources, where the theoretical curves with different parameters by RPM [AWOM] are also plotted. (c) [(d)] is similar to (a) [(b)], but only for the data of three sources Cyg X-2, SAX J1808.4-3658 and 4U 1820-30, together with the fitted model curves.}
\label{A_GR_model}
\end{figure*}
Kilohertz quasi-periodic oscillations (kHz QPOs) are the timing phenomena that exhibit the millisecond accretion flow close to
the surface of
 neutron star (NS) in low mass X-ray binaries (LMXBs, \citealt{Liu07}). They usually occur in pairs (upper $\nu_2$ and lower $\nu_1$) and cover the frequency range from tens of Hz to nearly 1300\,Hz \citep{van der Klis00,van der Klis06}. Until now there are 26 sources of exhibiting the twin kHz QPOs, including 16 Atoll, 8 Z sources, as well as 2 {\bf accreting } millisecond pulsars.
The twin kHz QPO frequencies follow the tight {\bf correlation}, showing a power-law relation \citep{Zhang06a}.
Furthermore, they also correlate with the low-frequency QPOs {\bf in a systematic way} \citep{Psaltis99,Belloni02}.

Many authors have investigated the correlation between $\nu_1$ and $\nu_2$: \citet{Stella99a} and \citet{Stella99b} propose the relativistic precession model (RPM) with the NS mass as the solely free parameter, where $\nu_2$ is the Keplerian orbital frequency and $\nu_1$ is the precession frequency of the orbit periastron. \citet{Torok08} apply this model to estimate the NS mass of Cir X-1 and give a value of $\sim2.2\,M_\odot$\citep{Torok10}. \citet{Zhang04} proposes the Alfv\'en wave oscillation model (AWOM) with the NS density as the free parameter, where $\nu_2$ is the Keplerian orbital frequency and $\nu_1$ is the MHD Alfv\'en wave oscillation frequency.
\citet{Stuchlik08,Stuchlik11} interpret the high frequency QPOs as the forced resonant oscillations of the accretion disk excited by gravitational perturbations. Other models, such as the beat frequency model \citep{Miller98,Lamb01} and constant ratio model \citep{Kluzniak01,Abramowicz03a,Abramowicz03b} are not strongly {\bf  supported} by the subsequent observations \citep{Mendez98,Mendez99,Jonker02,Wijnands03,Belloni05,Belloni07,Strohmayer06,Zhang06a,Yin07}.

The kHz QPO phenomenon is a powerful tool to probe the physical process around {\bf the surface of } NS \citep{Zhang13}: First, it can test Einstein's General Relativity {\bf in a strong gravitational  field } because its frequency range is near the dynamical timescale of the innermost region {\bf close to } NS surface  \citep{Kluzniak90,Lai98,van der Klis00,Mendez02,van der Klis06}.
Second, it can constrain the NS $Mass-Radius$ ($M-R$) relation  by the kHz QPO models \citep{Miller98,Stella99a,Stella99b,Zhang04}, {which will further} infer the matter state of NS with the help of equations of state (EOSs) \citep{Haensel07,Zhang07,Zhang09,Bhattacharyya10a,Petri11}.
So far, \citet{Kaaret02} {\bf has reported  a maximum frequency } of 1253\,Hz in  source SAX J1750.8-2900 (However, \citet{Bhattacharyya10b} reported a possible  1860\,Hz QPO in source 4U 1636-536, which may be the overtone of the kHz QPOs).

In this paper, we test the twin kHz QPO models of RPM and AWOM by  the sources (Cyg X-2, SAX J1808.4-3658 and 4U 1820-30) with measured NS masses and twin kHz QPOs, and constrain the NS $M-R$ relations and infer the possible matter compositions of these NSs.
Finally, we present the discussions and conclusions.

\section{Testing RPM and AWOM}

Both RPM and AWOM predict an approximate power law relation between $\nu_1$ and $\nu_2$ \citep{Stella99a,Stella99b,Zhang04},
then they are based on the different physical mechanisms and rely on the different parameters: RPM and AWOM rely on the NS mass and the NS density, repectively.
The two models assume that the upper-frequency of the twin kHz QPOs is the Keplerian orbital frequency at radius r:

\begin{equation}
\nu_2=\sqrt{\frac{GM}{4\pi^2r^3}}
\label{R_K}
\end{equation}
where $M$ is the NS mass.
RPM predicts the following relation between $\nu_1$ and $\nu_2$ \citep{Stella99a,Stella99b}\footnote{Geometrical units ($G=c=1$) are used in RPM, where we use  cgs units.}:

\begin{equation}
\Delta\nu=\nu_2\sqrt{1-\frac{3R_{\rm s}}{r}}
\label{GR_Fun1}
\end{equation}
where $R_{\rm s}$ is the Schwarzschild radius (i.e. $R_{\rm s}=2GM/{c^2}$).
Combined equation (\ref{R_K}) and equation (\ref{GR_Fun1}), the NS mass is the function of $\nu_1$ and $\nu_2$.
AWOM predicts the relation of $\nu_1$ and $\nu_2$ as follows \citep{Zhang04}:

\begin{equation}
\nu_1=629(\rm Hz)A^{-2/3}\nu_{2\rm k}^{5/3}\sqrt{1-\sqrt{1-(\frac{\nu_{2\rm k}}{1.85A})^{2/3}}}
\label{A_Fun1}
\end{equation}

\begin{equation}
A=\sqrt{\frac{m}{R^3_6}}
\label{A_Fun2}
\end{equation}
\begin{equation}
\rho=4.75\times10^{14}A^2({\rm g\,cm^{-3}})
\label{A_Fun3}
\end{equation}
\begin{table*}
\centering
\begin{minipage}{160mm}
\centering
\caption{Sources with the measured NS masses and twin kHz QPOs}
\begin{tabular}{@{}lccccccr@{}}
\hline
%\centering
Source & Measured Mass & Ref & $R_{\rm ISCO}$ & $\nu_1$ & $\nu_2$ & Ref & $R_{\rm K}$ \\
%%\hline
(3) & ($M_\odot$) & & (km) & (Hz) & (Hz) & & (km) \\
\hline
\multirow{2}{*}{Cyg X-2} & $1.78\pm0.23$ & 1 & $15.8\pm2.0$ & \multirow{2}*{$516\pm27$} & \multirow{2}*{$862\pm11$} & \multirow{2}*{5} & $20.0\pm0.9$ \\
 & $1.5\pm0.3$ & 2 & $13\pm3$ & & & & $19\pm1$ \\
SAX J1808.4-3658 & $<1.4$ & 3 & $<12$ & $499\pm4\sim503.6\pm5.3$ & $685.1\pm5.1\sim694\pm4$ & 6 & $<21\pm0.1$ \\
4U 1820-30 & $1.29^{+0.19}_{-0.07}$ & 4 & $11.4^{+1.7}_{-0.6}$ & $764\pm6\sim796\pm4$ & $1055\pm10\sim1072\pm11$ & 7 & $15.6^{+0.8}_{-0.3}$ \\
\hline
\end{tabular}
\end{minipage}
\begin{tabular}{@{}l@{}}
%\centering
\begin{minipage}{160mm}
Notes: The second and forth columns show the measured NS masses and the inferred innermost stable circular orbit (ISCO) radii (i.e. $R_{\rm ISCO}=3R_{\rm s}=6GM/c^2$), the fifth and sixth columns show the range of $\nu_1$ and $\nu_2$, the last column shows the Keplerian orbital radii inferred from the maxima of $\nu_2$.
References:
1. \citealt{Orosz99};
2. \citealt{Elebert09b};
3. \citealt{Elebert09a}; %\citealt{Chakrabarty98},
4. \citealt{Shaposhnikov04}; %\citealt{Wang Z10},
5. \citealt{Wijnands98};
6. \citealt{van Straaten05}, \citealt{Wijnands03};
7. \citealt{Smale97}
\end{minipage}
\end{tabular}
\label{mass}
\end{table*}
\begin{table*}
\begin{minipage}{115mm}
\centering
\caption{The derived parameters by RPM and AWOM of the sources in Table 1}
\begin{tabular}{@{}lccccc@{}}
\hline
%\centering
Source & m (RPM) & $\chi^2/d.o.f.$ & A (AWOM) & $\chi^2/d.o.f.$ & Radius (AWOM) \\
%%\hline
(3) & ($M_\odot$) & & & & ($\rm km$) \\ %($\propto\,M_\odot/10km^3$)
\hline
\multirow{2}{*}{Cyg X-2} & \multirow{2}*{$1.96\pm0.10$} & \multirow{2}*{--} & \multirow{2}*{$0.62\pm0.04$} & \multirow{2}*{--} & $16.7\pm1.1$ \\
 & & & & & $16\pm1$ \\
SAX J1808.4-3658 & $2.83\pm0.04$ & -- & $0.43\pm0.01$ & --  & $<20$ \\
4U 1820-30 & $1.85\pm0.02$ &  2.76 & $0.65\pm0.01$ & 2.58 & $14.5^{+0.7}_{-0.3}$ \\
\hline
\end{tabular}
\end{minipage}
\begin{tabular}{@{}l@{}}
\begin{minipage}{114mm}
Notes: The second and third (forth and fifth) columns show the inferred NS masses (density parameter $A$ values) and the goodness of fit by RPM (AWOM), respectively. The source Cyg X-2 and SAX J1808.4-3658 have too few data to fit, therefore we have to calculate the mean parameters for the models. The sixth column shows the inferred NS radii based on $A$ values (AWOM) and measured NS masses in Table 1.
\end{minipage}
\end{tabular}
\label{model}
\end{table*}
\begin{figure*}
\includegraphics[width=7.5cm]{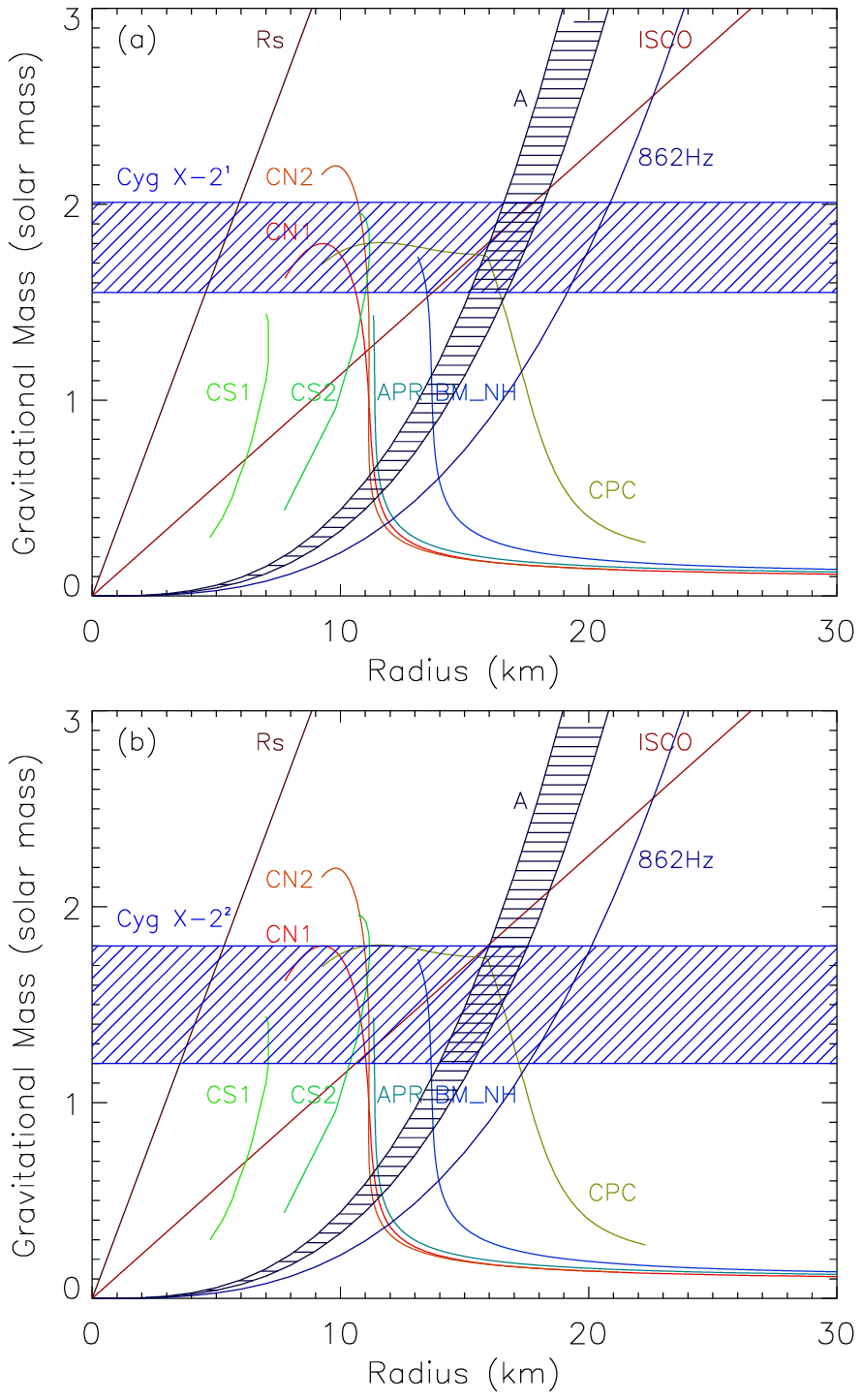}
\includegraphics[width=7.5cm]{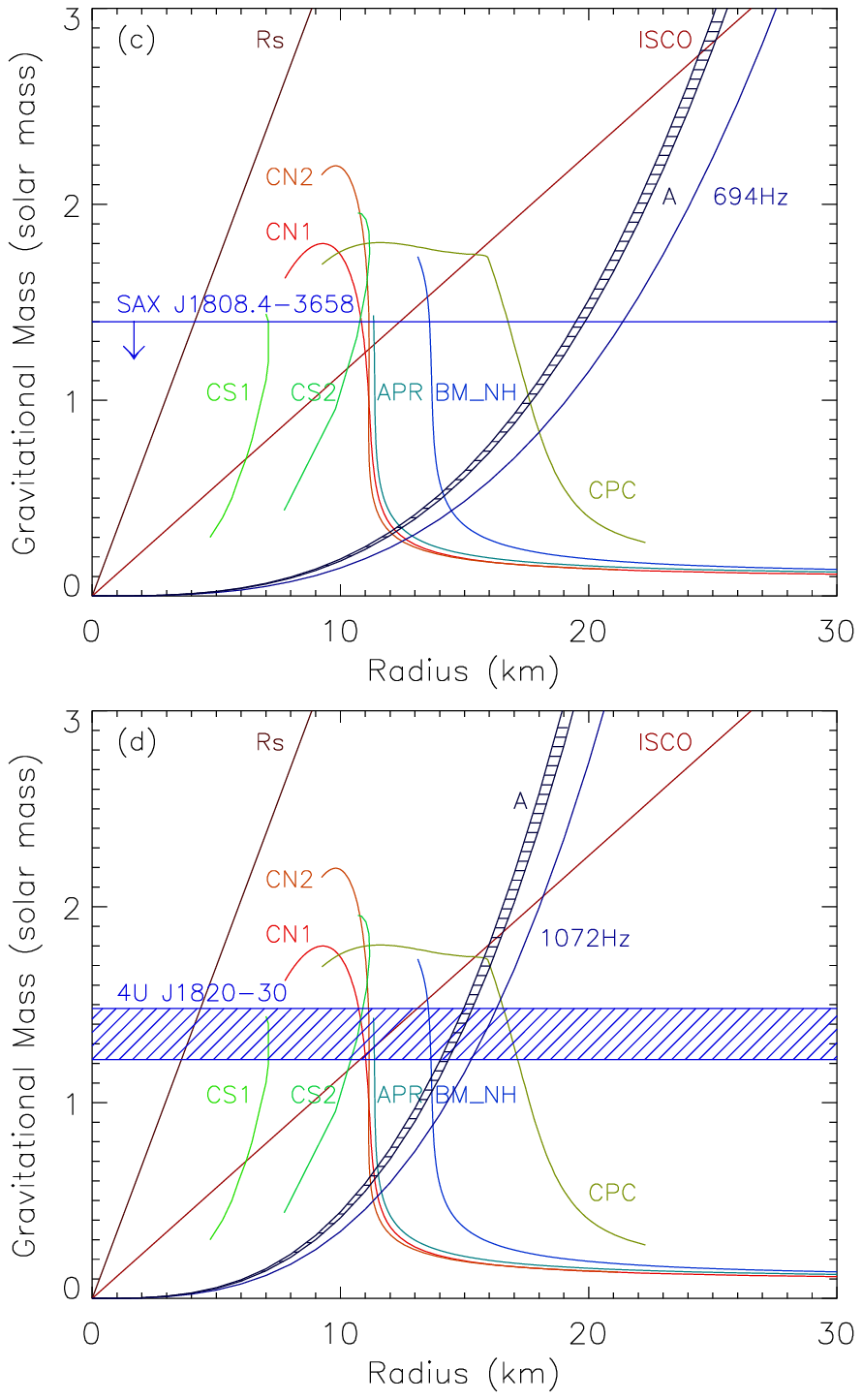}
\caption{
The $M-R$ diagram of NS.
Some representative EOS curves are shown (see also e.g. \citealt{Akmal98,Miller02,Bednarek12}):
stars containing strange quark matter (CS1 and CS2),
nucleons only (APR) \citep{Akmal98},
nucleons and hyperons (BM\_NH) \citep{Bednarek12},
stars made of normal neutron matter (CN1 and CN2)
and with pion condensate cores (CPC) \citep{Lattimer01,Cook94}.
$R_{\rm s}$ represents the Schwarzschild radius, and
ISCO represents the innermost stable circular orbit.
For each source in Table 1, we show the range of measured NS mass (see Table 1) and $M-R$ range based on $A$ value (see Table 2).
(a)-(b) is for Cyg X-2 (with the NS mass $1.78\pm0.23\,M_\odot$ and $1.5\pm0.3\,M_\odot$, respectively), (c) is for SAX J1808.4-3658 and (d) is for 4U 1820-30.
}
\label{m_r_test}
\end{figure*}
where $\nu_{2\rm k}=\nu_2/1000\rm \,Hz$, $m$ is the NS mass in $M_\odot$,  $R_6=R/10^6\rm \,cm$ is the NS radius $R$ in 10\,km and $A$ is a parameter that presents  the measurement of the NS density $\rho$.

Since the NS mass is the free parameter of RPM, we can test this model by comparing the predicted NS mass by twin kHz QPOs with the measured result. While for AWOM, the free parameter is NS density, we can infer the NS radius from the twin kHz QPOs and measured NS mass, by which we can justify if this radius is a reasonable value. In summary, in order to test  the two models, we need the sources to have both the measured NS masses and twin kHz QPOs. We find only three sources (Cyg X-2, SAX J1808.4-3658 and 4U 1820-30) to satisfy {\bf these conditions}, as shown in Table 1, which are selected from the updated 393 pairs of twin kHz QPOs (see, e.g. \citealt{van der Klis06,Zhang06a,Linares05,Boutloukos06,Zhang06b,Homan07,Altamirano08,Barret08,Strohmayer08,Boutelier09,Altamirano10,Bhattacharyya10b,Homan10,Sanna10,Lin11}).
The particular properties of these three sources are briefly summarized below:

Cyg X-2 is a persistent NS-LMXB source with the luminosity close to the Eddington limit. It locates at a distance of $\sim7$ kpc \citep{Orosz99} with the orbital period of 9.8444\,d \citep{Casares98} and the secondary of $\sim0.6\,M_\odot$ \citep{Podsiadlowsk00}. Both type I X-ray bursts \citep{Smale98} and twin kHz QPOs \citep{Wijnands98} have been observed in this source.  By analyzing the high-resolution optical spectroscopy and the rotationally broadened absorption features of the secondary star, \citet{Casares98} determine the optical mass function of $f(M)=0.69\pm0.03\,M_\odot$ and the mass ratio of $q=M_c/M_x=0.34\pm0.04$, based on which and adopting the inclination $i=62.5^\circ\pm4^\circ$, \citet{Orosz99} find that the NS mass in this system is
$M_x=1.78\pm0.23\,M_\odot$. Besides this measurement, \citet{Elebert09b} analyze the radial velocity (RV) curve for the secondary star, derive the mass function and further imply the NS mass {\bf to be } $1.5\pm0.3\,M_\odot$.

SAX J1808.4-3658 contains a neutron star with a very low-mass companion star in a 2-h
orbital period \citep{Chakrabarty98}, the estimated distance is 2.5 kpc \citep{In't Zand98,In't Zand01} and the maximum luminosity is $\sim 10^{36}\,\rm erg~s^{-1}$. This source has shown type-I X-ray bursts \citep{In't Zand98,In't Zand01,Chakrabarty03}, coherent 401 Hz pulsation \citep{Wijnands98a}, and the  simultaneous twin kHz QPOs \citep{Wijnands03}. Based on the existing pulse timing measurements, \citet{Elebert09a} constrain the mass ratio of the system and infer the mass function for the pulsar to be 0.44\,$M_\odot$, from which the NS mass of this source is derived to be less than 1.4\,$M_\odot$. \\

4U 1820-30 is a NS-LMXB source residing in the globular cluster NGC 6624. It has the extremely short
 orbital period of 685\,s and its secondary is a low-mass helium-rich degenerate star \citep{Stella87}.
This source has an estimated distance of $6.4\pm0.6$\,kpc \citep{Vacca86} from UV observation and 7.6 kpc from optical
observations \citep{Rich93}, and  has been detected {\bf with} both Type I X-ray bursts  \citep{Grindlay76} and twin kHz QPOs \citep{Smale97}. Based on the observed properties of X-ray bursts, \citet{Shaposhnikov04} infer the mass-radius relationship by modeling the spectral temperature dependence
on the bolometric flux, and further infer the NS mass of $1.29^{+0.19}_{-0.07}\,M_\odot$  and radius of $11.2^{+0.4}_{-0.5}$\,km.

We adopt these three sources  to test the twin kHz QPO model RPM and AWOM, {\bf and }the inferred parameters are shown in Table 2. As for RPM, the derived NS mass of Cyg X-2 (SAX J1808.4-3658 and 4U 1820-30) by twin kHz QPOs is $1.96\pm0.10\,M_\odot$ ($2.83\pm0.04\,M_\odot$ and $1.85\pm0.02\,M_\odot$), while the corresponding measured result is $1.5\pm0.3\,M_\odot$ ($<1.4\,M_\odot$ and $1.29^{+0.19}_{-0.07}\,M_\odot$). It can be seen that all the inferred NS masses by RPM are systemically bigger than the measured results by $\sim50\,\%$, especially for SAX J1808.4-3658, whose predicted NS mass is twice of the measured result. As for AWOM, we derive the values of the density parameter $A$  of  three sources by twin kHz QPOs, then infer the NS radii as $10-20$\,km by the measured NS masses, which satisfy the theoretical  expectations of NS EOSs.

Fig.\ref{A_GR_model} (a) [(b)] shows the plot of $\Delta\nu~vs.~\nu_2$ with the twin kHz QPOs and the curves by RPM [AWOM]. It can be seen that most
 {\bf data points of} twin kHz QPOs center at the region with the parameters $m=2\,M_\odot$ for RPM and $A=0.6$ for AWOM. The data positions of the {\bf accreting} millisecond pulsars XTE J1807-294 and SAX J1808.4-3658 in Fig.\ref{A_GR_model} are systemically lower than those of the other sources. Especially, for XTE J1807-294, the corresponding  parameter of RPM is $m=3.2\,M_\odot$, which is the mass limit of NS introduced by \citet{Rhoades74}. Therefore, we can conclude that the RPM model overestimates the NS mass.

\section{M-R constrain of NS}

$M-R$ relation is helpful for understanding the nuclear compositions inside NS, and many methods are proposed to constrain it, such as by rotational broadening, redshift, Eddington limit, surface emission \citep{Ozel06,Zhang07} and kHz QPOs \citep{Miller98}. In addition, \citet{Zhang09} adopts AWOM to constrain $M-R$ relation by the twin kHz QPOs.

With the twin kHz QPOs and equation (\ref{A_Fun2}), we {\bf firstly} calculate the {\bf  density parameter}  $A$ values of the sources  Cyg X-2, SAX J1808.4-3658 and 4U 1820-30, then constrain their NS radii with the measured NS masses. By comparing the measured NS mass and derived radius  with the EOS curves {\bf in M-R diagram}, we can infer the possible matter compositions inside these NSs, which are shown in Fig.\ref{m_r_test}. For Cyg X-2, its NS is likely to contain the condensate pion core (CPC); for SAX J1808.4-3658,  the  EOS of NS may exclude strange quark matter (CS1 and CS2); for 4U 1820-30, its NS may contain the condensate pion core, or nucleons and hyperons (BM\_NH).

\begin{figure}
\includegraphics[width=8cm]{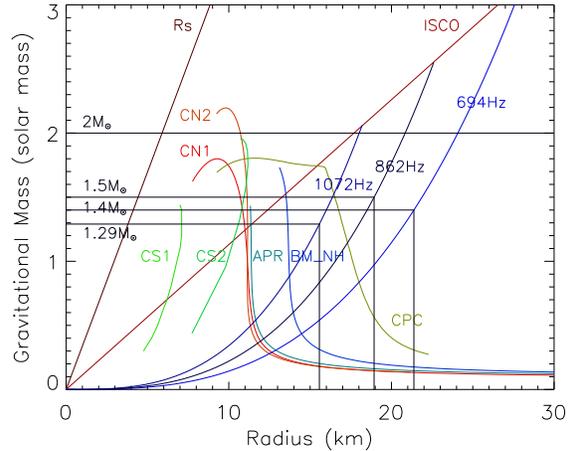}
\caption{
The $M-R$ diagram of NS, where the EOS curves are same as those in Fig.\ref{m_r_test}.
The curves correspond to the maxima of $\nu_2$  of  Cyg X-2 (862 Hz), SAX J1808.4-3658 (694 Hz) and 4U 1820-30 (1072 Hz), which are assumed as the Keplerian orbital frequencies, and the measured NS masses and derived NS radius ranges are also plotted.}
\label{max_frequency}
\end{figure}

Furthermore, we constrain the $M-R$ relations by the maxima of kHz QPOs, assumed to be the Keplerian orbital frequencies, for Cyg X-2, SAX J1808.4-3658 and 4U 1820-30, which are shown in Fig.\ref{max_frequency}. The constrained radii are listed in the last column of Table 1.   With the certain NS mass, the EOS of NS is constrained between the line of Schwarzschild radius $R_{\rm s}$ and the curve where Keplerian orbital frequency is the maxima of kHz QPOs, then this method cannot present a definite conclusion for NS EOSs. Most EOSs shown in the figure are compatible for all three sources.
It can be also seen that among the EOSs shown in Fig.\ref{max_frequency}, stars made of normal neutron matter (CN2) can sustain a NS of 2 $M_\odot$.

\section{Discussions and Conclusions}

By exploiting the three sources, Cyg X-2, SAX J1808.4-3658 and 4U 1820-30, with the measured NS masses and twin kHz QPOs, we  test the models of twin kHz QPOs (RPM and AWOM), and constrain their NS EOSs. The summaries and discussions are listed below:

(1). We test the twin kHz QPO models RPM and AWOM by the sources with the  measured NS masses and twin kHz QPOs. From the results in Table 2, it can be seen that the inferred NS masses by RPM are all systemically bigger than the measured results by $\sim50\,\%$. In particular, the inferred NS mass ($2.83\pm0.04$) of SAX J1808.4-3658 is twice of the measured result ($<1.4\,M_\odot$). It can be seen from Fig.\ref{A_GR_model} that most data points of the twin kHz QPOs center at the predicted curve by RPM with NS mass of $2\,M_\odot$.
RPM overestimates the NS mass of Cyg X-2 ($1.96\pm0.10\,M_\odot$) by $\sim30\,\%$ than the measured result ($1.5\pm0.3\,M_\odot$). We take this scale factor as the reference and infer
the mean value of NS mass to be $1.54\,M_\odot$, which is close to the statistical result by \citet{Zhang11} ($1.57\pm0.35\,M_\odot$ for the measured NS masses of MSPs).
The predicted NS mass by RPM for XTE 1807-294 is near $3.2\,M_\odot$ (see also Fig.\ref{A_GR_model}). The twin kHz QPO properties  of XTE 1807-294 and SAX J1808.4-3658 are similar, so we take the scale factor $\sim100\,\%$ to infer the NS mass of XTE 1807-294 to be $1.6\,M_\odot$. In the  construction of RPM, only the ideal environments are considered, such as vacuum condition and test particle, which is too simplified  to satisfy the reality of the accreting NS, and we think that the modification of this model is necessary by considering the interaction between the NS and accretion disk, the effect of the magnetic field.
The inferred NS radii of the three sources (Cyg X-2, SAX J1808.4-3658 and 4U 1820-30) by AWOM are around $10-20\,\rm km$ (see Table 2), which satisfy the theoretical expectations of NS EOSs.

(2).
We constrain the NS $M-R$ relations of the sources in Table 1 by twin kHz QPOs and AWOM, and infer the possible matter compositions of these NSs with the help of EOSs,
which are  shown in Fig.\ref{m_r_test}. The measured NS mass of SAX J1808.4-3658 is $<1.4\,M_\odot$ with the inferred density parameter $A=0.43\pm0.01$ by AWOM, which is less than the values of the other sources with $A\sim0.6$. The EOS of this NS may exclude the strange quark matter (see Fig.\ref{m_r_test}, comparing with the relative reference by \citealt{Li99}). By adopting the NS mass range of $1-1.4\,M_\odot$ ($1\,M_\odot$ is the measured lower limit of the NS mass, see \citealt{Miller02,Zhang11}) and $A$ value, we infer its NS radius range to be $17-20\,\rm km$, and induce that the EOS of its NS is close to the state with condensate pion core (CPC). Similarly, the EOS of Cyg X-2 is also near CPC.

For 4U 1820-30, \citet{Shaposhnikov04} infer its NS mass of $1.29^{+0.19}_{-0.07}\,M_\odot$ and radius of $11.2^{+0.4}_{-0.5}$\,km according to the observed properties of X-ray bursts. The authors adopt the photospheric radius expansion model to fit the NS mass, radius and some other parameters simultaneously, in which some theoretical assumptions are set, such as spectral formation, dynamic evolution of NS-disk geometry and the helium abundance $Y_{\rm He}$ (to be 1.0). Based on AWOM, we infer the NS density  parameter $A$ ($0.65\pm0.01$) of 4U 1820-30, where two theoretical assumptions (Keplerian orbital frequency for $\nu_2$ and MHD Alfv\'en wave oscillation frequency for $\nu_1$) are employed. Furthermore, the NS radius of this source is inferred to be $14.5^{+0.7}_{-0.3}$\,km by adopting the NS mass of $1.29^{+0.19}_{-0.07}\,M_\odot$ \citep{Shaposhnikov04}, which is bigger than their result ($11.2^{+0.4}_{-0.5}$\,km). Considering the model dependence of the two results, we cannot make a certain conclusion which radius is more reasonable.

\section*{Acknowledgments}

This work is supported by National Basic Research Program of China
(2012CB821800 and 2009CB824800),  National Natural Science
Foundation of China NSFC(11173034, 11173024, 10773017, 10778716,
11203064, 10903005) and Fundamental
Research Funds for the Central Universities.

\bsp

\label{lastpage}


\begin{thebibliography}{}

\bibitem[\protect\citeauthoryear{Abramowicz et al.}{2003a}]{Abramowicz03a}
Abramowicz M. A. et al., 2003a, A\&A, 404, L21

\bibitem[\protect\citeauthoryear{Abramowicz et al.}{2003b}]{Abramowicz03b}
Abramowicz M. A. et al., 2003b, PASJ, 55, 467

\bibitem[\protect\citeauthoryear{Akmal et al.}{1998}]{Akmal98}
Akmal A., Pandharipande V. R., Ravenhall D. G., 1998, Phys. Rev. C, 58, 1804

\bibitem[\protect\citeauthoryear{Altamirano et al.}{2008}]{Altamirano08}
Altamirano D. et al., 2008, ApJ, 685, 436
%
\bibitem[\protect\citeauthoryear{Altamirano et al.}{2010}]{Altamirano10}
Altamirano D. et al., 2010, MNRAS, 401, 223
%
\bibitem[\protect\citeauthoryear{Barret et al.}{2008}]{Barret08}
Barret D., Boutelier M., Miller M. C., 2008, MNRAS, 384, 1519
%
\bibitem[\protect\citeauthoryear{Bednarek et al.}{2012}]{Bednarek12}
Bednarek I. et al., 2012, A\&A, 543, A157
%
\bibitem[\protect\citeauthoryear{Belloni et al.}{2002}]{Belloni02}
Belloni T., Psaltis D., van der Klis, M., 2002, ApJ, 572, 392
%
\bibitem[\protect\citeauthoryear{Belloni et al.}{2005}]{Belloni05}
Belloni T., M\'endez M., Homan J., 2005, A\&A, 437, 209
%
\bibitem[\protect\citeauthoryear{Belloni et al.}{2007}]{Belloni07}
Belloni T., M\'endez M., Homan J., 2007, MNRAS 376, 1133
%
\bibitem[\protect\citeauthoryear{Bhattacharyya}{2010a}]{Bhattacharyya10a}
Bhattacharyya S., 2010a, Advances in Space Research, 45, 949
%
\bibitem[\protect\citeauthoryear{Bhattacharyya}{2010b}]{Bhattacharyya10b}
Bhattacharyya S., 2010b, Research in Astronomy and Astrophysics, 10, 227
%
\bibitem[\protect\citeauthoryear{Boutelier et al.}{2009}]{Boutelier09}
Boutelier M., Barret D., Miller M. C., 2009, MNRAS, 399, 1901
%
\bibitem[\protect\citeauthoryear{Boutloukos et al.}{2006}]{Boutloukos06}
Boutloukos S. et al., 2006, ApJ, 653, 1435
%
\bibitem[\protect\citeauthoryear{Casares et al.}{1998}]{Casares98}
Casares J., Charles P. A., Kuulkers E., 1998, ApJ, 493, L39
%
\bibitem[\protect\citeauthoryear{Chakrabarty \& Morgan}{1998}]{Chakrabarty98}
Chakrabarty D., Morgan E. H., 1998, Nat, 394, 346
%
\bibitem[\protect\citeauthoryear{Chakrabarty et al.}{2003}]{Chakrabarty03}
Chakrabarty D. et al., 2003, Nat, 424, 42
%
\bibitem[\protect\citeauthoryear{Cook et al.}{1994}]{Cook94}
Cook G. B., Shapiro S. L., Teukolsky S. A., 1994, ApJ, 424, 823
%
\bibitem[\protect\citeauthoryear{Elebert et al.}{2009a}]{Elebert09a}
Elebert P. et al., 2009a, MNRAS, 395, 884
%
\bibitem[\protect\citeauthoryear{Elebert et al.}{2009b}]{Elebert09b}
Elebert P. et al., 2009b, MNRAS, 395, 2029
%
\bibitem[\protect\citeauthoryear{Grindlay et al.}{1976}]{Grindlay76}
Grindlay J. E., Gursky H., Schnopper H., 1976, ApJ, 205, L127
%
\bibitem[\protect\citeauthoryear{Haensel et al.}{2007}]{Haensel07}
Haensel P., Potekhin A. Y., Yakovlev D. G., 2007, Neutron Stars 1: Equation
of State and Structure. Springer, New York
%
\bibitem[\protect\citeauthoryear{Homan et al.}{2007}]{Homan07}
Homan J. et al., 2007, ApJ, 656, 420
%
\bibitem[\protect\citeauthoryear{Homan et al.}{2010}]{Homan10}
Homan J. et al., 2010, ApJ, 719, 201
%
\bibitem[\protect\citeauthoryear{In't Zand et al.}{1998}]{In't Zand98}
In't Zand J. J. M. et al., 1998, A\&A, 331, L25
%
\bibitem[\protect\citeauthoryear{In't Zand et al.}{2001}]{In't Zand01}
In't Zand J. J. M. et al, 2001, A\&A, 372, 919
%
\bibitem[\protect\citeauthoryear{Jonker et al.}{2002}]{Jonker02}
Jonker P. G., M\'endez M., van der Klis M., 2002, MNRAS, 336, L1
%
\bibitem[\protect\citeauthoryear{Kaaret et al.}{2002}]{Kaaret02}
Kaaret P. et al., 2002, ApJ, 575, 1018
%
\bibitem[\protect\citeauthoryear{Klu\'zniak et al.}{1990}]{Kluzniak90}
Klu\'zniak W., Michelson P., Wagoner R. V., 1990, ApJ, 358, 538
%
\bibitem[\protect\citeauthoryear{Klu\'zniak \& Abramowicz}{2001}]{Kluzniak01}
Klu\'zniak W., Abramowicz M. A., 2001, Acta Phys. Pol. B, 32, 3605
%
\bibitem[\protect\citeauthoryear{Lai}{1998}]{Lai98}
Lai D., 1998, ApJ, 502, 721
%
\bibitem[\protect\citeauthoryear{Lamb \& Miller}{2001}]{Lamb01}
Lamb F. K., Miller M. C. 2001, ApJ, 554, 1210
%
\bibitem[\protect\citeauthoryear{Lattimer \& Prakash}{2001}]{Lattimer01}
Lattimer J. M., Prakash M., 2001, ApJ, 550, 426
%
\bibitem[\protect\citeauthoryear{Li et al.}{1999}]{Li99}
Li X. D. et al., 1999, Phys. Rev. Lett., 83, 3776
%
\bibitem[\protect\citeauthoryear{Lin et al.}{2011}]{Lin11}
Lin Y. F. et al., 2011, ApJ, 726, 74
%
\bibitem[\protect\citeauthoryear{Linares et al.}{2005}]{Linares05}
Linares M. et al., 2005, ApJ, 634, 1250
%
\bibitem[\protect\citeauthoryear{Liu et al.}{2007}]{Liu07}
Liu Q. Z. et al., 2007, A\&A, 469, 807
%
\bibitem[\protect\citeauthoryear{M\'endez et al.}{1998}]{Mendez98}
M\'endez M. et al., 1998, ApJ, 505, L23
%
\bibitem[\protect\citeauthoryear{M\'endez \& van der Klis}{1999}]{Mendez99}
M\'endez M., van der Klis M., 1999, ApJ, 517, L51
%%sco X-1
%
\bibitem[\protect\citeauthoryear{M\'endez}{2002}]{Mendez02}
M\'endez M., 2002, in Gurzadyan V. G., Jantzen R. T., Ruffini R., eds., The Ninth Marcel Grossmann Meeting, Proceedings of the MGIX MM Meeting, World Scientific Publishing, Co. Pte. Ltd., in 3 volumes, p. 2319
%
\bibitem[\protect\citeauthoryear{Miller et al.}{1998}]{Miller98}
Miller M. C., Lamb F. K., Psaltis D., 1998, ApJ,  508, 791
%
\bibitem[\protect\citeauthoryear{Miller}{2002}]{Miller02}
Miller M. C., 2002, Nat, 420, 31
%
\bibitem[\protect\citeauthoryear{Orosz \& Kuulkers}{1999}]{Orosz99}
Orosz J. A., Kuulkers E., 1999, MNRAS, 305, 132
%
\bibitem[\protect\citeauthoryear{\"Ozel}{2006}]{Ozel06}
\"Ozel F., 2006, Nat, 441, 1115
%
\bibitem[\protect\citeauthoryear{P\'etri}{2011}]{Petri11}
P\'etri J., Ap\&SS, 2011, 331, 555
%
\bibitem[\protect\citeauthoryear{Podsiadlowski \& Rappaport}{2000}]{Podsiadlowsk00}
Podsiadlowski P., Rappaport S., 2000, ApJ, 529, 946
%
\bibitem[\protect\citeauthoryear{Psaltis et al.}{1999}]{Psaltis99}
Psaltis D., Belloni T., van der Klis M., 1999, ApJ, 520, 262 %(PBK99)
%
\bibitem[\protect\citeauthoryear{Rhoades \& Ruffini}{1974}]{Rhoades74}
Rhoades C. E., Ruffini R., 1974, Phys. Rev. Lett., 32, 324
%
\bibitem[\protect\citeauthoryear{Rich et al.}{1993}]{Rich93}
Rich R. M., Minniti D., Liebert J., 1993, ApJ, 406, 489
%
\bibitem[\protect\citeauthoryear{Sanna et al.}{2010}]{Sanna10}
Sanna A. et al., 2010, MNRAS, 408, 622
%
\bibitem[\protect\citeauthoryear{Smale et al.}{1997}]{Smale97}
Smale A. P., Zhang W., White N. E., 1997, ApJ, 483, L119
%
\bibitem[\protect\citeauthoryear{Smale}{1998}]{Smale98}
Smale A. P., 1998, ApJ, 498, L141
%
\bibitem[\protect\citeauthoryear{Shaposhnikov \& Titarchuk}{2004}]{Shaposhnikov04}
Shaposhnikov N., Titarchuk L., 2004, ApJ, 606, L57
%
\bibitem[\protect\citeauthoryear{Stella et al.}{1987}]{Stella87}
Stella L., Priedhorsky W., White N. E., 1987, ApJ, 312, L17
%
\bibitem[\protect\citeauthoryear{Stella \& Vietri}{1999}]{Stella99a}
Stella L., Vietri M., 1999, Phys. Rev. Lett., 82, 17 %(SV99)
%
\bibitem[\protect\citeauthoryear{Stella et al.}{1999}]{Stella99b}
Stella L., Vietri M., Morsink S. M., 1999, ApJ, 524, L63
%
\bibitem[\protect\citeauthoryear{Strohmayer \& Bildsten}{2006}]{Strohmayer06}
Strohmayer T. E., Bildsten L., 2006, in Lewin W. H. G., van der Klis M., eds,
Compact Stellar X-Ray Sources, Cambridge Univ. Press, Cambridge, p.113
%
\bibitem[\protect\citeauthoryear{Strohmayer et al.}{2008}]{Strohmayer08}
Strohmayer T. E., Markwardt C. B., Swank J. H., 2008, Astron. Telegram, 1635, 1
%
\bibitem[\protect\citeauthoryear{Stuchl\'ik et al.}{2008}]{Stuchlik08}
Stuchl\'ik Z. et al., 2008, A\&A, 489, 963
%
\bibitem[\protect\citeauthoryear{Stuchl\'ik et al.}{2011}]{Stuchlik11}
Stuchl\'ik Z., Kotrlov\'a A., T\"or\"ok G., 2011, A\&A, 525, A82
%
\bibitem[\protect\citeauthoryear{T\"or\"ok et al.}{2008}]{Torok08}
T\"or\"ok G. et al., 2008, Acta Astron., 58, 1
%
\bibitem[\protect\citeauthoryear{T\"or\"ok et al.}{2010}]{Torok10}
T\"or\"ok G. et al., 2010, ApJ, 714, 748
%
\bibitem[\protect\citeauthoryear{Vacca et al.}{1986}]{Vacca86}
Vacca W. D., Lewin W. H. G., van Paradijs J., 1986, MNRAS, 220, 339
%
\bibitem[\protect\citeauthoryear{van der Klis}{2000}]{van der Klis00}
van der Klis M., 2000, ARA\&A, 38, 717 %[arxiv;astro-ph/0001167]
%
\bibitem[\protect\citeauthoryear{van der Klis}{2006}]{van der Klis06}
van der Klis M., 2006, in Lewin W. H. G., van der Klis M., eds,
Compact Stellar X-Ray Sources, Cambridge Univ. Press, Cambridge, p.39
%
\bibitem[\protect\citeauthoryear{van Straaten et al.}{2005}]{van Straaten05}
van Straaten S., van der Klis M., Wijnands R., 2005, ApJ, 619, 455 % 1608 term scheme
%
\bibitem[\protect\citeauthoryear{Wijnands \& van der Klis}{1998}]{Wijnands98a}
Wijnands R., van der Klis M., 1998, Nat, 394, 344
%
\bibitem[\protect\citeauthoryear{Wijnands et al.}{1998}]{Wijnands98}
Wijnands R. et al., 1998, ApJ, 493, L87
%
\bibitem[\protect\citeauthoryear{Wijnands et al.}{2003}]{Wijnands03}
Wijnands R. et al., 2003, Nat, 424, 44
%
\bibitem[\protect\citeauthoryear{Yin et al.}{2007}]{Yin07}
Yin H. X. et al., 2007, A\&A, 471, 381
%
\bibitem[\protect\citeauthoryear{Zhang}{2004}]{Zhang04}
Zhang C. M., 2004, A\&A, 423, 401
%
\bibitem[\protect\citeauthoryear{Zhang et al.}{2006a}]{Zhang06a}
Zhang C. M. et al., 2006a, MNRAS, 366, 1373
%
\bibitem[\protect\citeauthoryear{Zhang et al.}{2007}]{Zhang07}
Zhang C. M. et al., 2007, Astron. Nachr., 328, 491
%
\bibitem[\protect\citeauthoryear{Zhang}{2009}]{Zhang09}
Zhang C. M., 2009, Astron. Nachr., 330, 398
%
\bibitem[\protect\citeauthoryear{Zhang et al.}{2011}]{Zhang11}
Zhang C. M. et al., 2011, A\&A, 527, 83

\bibitem[\protect\citeauthoryear{Zhang \& Wang}{2013}]{Zhang13}
Zhang C. M., Wang D. H., 2013, in Zhang C. M., Belloni T., M\'endez M. et al., eds, Feeding Compact Objects: Accretion on All Scales, Proceedings of the International Astronomical Union, IAU Symp. 290, Cambridge: Cambridge University Press, pp. 381-385

\bibitem[\protect\citeauthoryear{Zhang et al.}{2006b}]{Zhang06b}
Zhang F. et al., 2006b, ApJ, 646, 1116

\end{thebibliography}
\end{document}